\documentstyle[preprint,eqsecnum,prd,aps,epsf]{revtex}   % Preprint format
\begin{document}
\draft
\tightenlines
\preprint{FERMILAB-Pub-00/054-E}
\title{Inclusive production of $\omega$ mesons at large transverse momenta \\
in $\pi^-$Be interactions at 515\ GeV/$c$}
\author{ 
L.~Apanasevich,$^{4}$
J.~Bacigalupi,$^{1}$
W.~Baker,$^{3}$
M.~Begel,$^{9}$
S.~Blusk,$^{8}$
C.~Bromberg,$^{4}$
P.~Chang,$^{5}$
B.~Choudhary,$^{2}$
W.~H.~Chung,$^{8}$
L.~de~Barbaro,$^{9}$
W.~DeSoi,$^{9}$
W.~D\l ugosz,$^{5}$
J.~Dunlea,$^{9}$
E.~Engels,~Jr.,$^{8}$
G.~Fanourakis,$^{9}$
T.~Ferbel,$^{9}$
J.~Ftacnik,$^{9}$
D.~Garelick,$^{5}$
G.~Ginther,$^{9}$
M.~Glaubman,$^{5}$
P.~Gutierrez,$^{6}$
K.~Hartman,$^{7}$
J.~Huston,$^{4}$
C.~Johnstone,$^{3}$
V.~Kapoor,$^{2}$
J.~Kuehler,$^{6}$
C.~Lirakis,$^{5}$
F.~Lobkowicz,$^{9,\ddag}$
P.~Lukens,$^{3}$
S.~Mani,$^{1}$
J.~Mansour,$^{9}$
A.~Maul,$^{4}$
R.~Miller,$^{4}$
B.~Y.~Oh,$^{7}$
G.~Osborne,$^{9}$
D.~Pellett,$^{1}$
E.~Prebys,$^{9}$
R.~Roser,$^{9}$
P.~Shepard,$^{8}$
R.~Shivpuri,$^{2}$
D.~Skow,$^{3}$
P.~Slattery,$^{9}$
L.~Sorrell,$^{4}$
D.~Striley,$^{5}$
W.~Toothacker,$^{7}$
N.~Varelas,$^{9}$
D.~Weerasundara,$^{8}$
J.~J.~Whitmore,$^{7}$
T.~Yasuda,$^{5}$
C.~Yosef,$^{4}$
M.~Zieli\'{n}ski,$^{9}$
V.~Zutshi$^{2}$
\\
{~}\\
\centerline{(Fermilab E706 Collaboration)}
{~}\\
}
\address{
\centerline{$^{1}$University of California-Davis, Davis, California 95616}
\centerline{$^{2}$University of Delhi, Delhi, India 110007}
\centerline{$^{3}$Fermi National Accelerator Laboratory, Batavia,              
                   Illinois 60510}
\centerline{$^{4}$Michigan State University, East Lansing, Michigan 48824}     
\centerline{$^{5}$Northeastern University, Boston, Massachusetts  02115}
\centerline{$^{6}$University of Oklahoma, Norman, Oklahoma  73019}
\centerline{$^{7}$Pennsylvania State University, University Park, 
		   Pennsylvania 16802}
\centerline{$^{8}$University of Pittsburgh, Pittsburgh, Pennsylvania 15260}
\centerline{$^{9}$University of Rochester, Rochester, New York 14627}          
\centerline{$^{\ddag}$Deceased}
}

\date{\today}
\maketitle
\begin{abstract}
We present results on the production of $\omega$ mesons at high
transverse momenta in $\pi^-$Be interactions at $515$~GeV/$c$.  The
data span the kinematic range $3.5 \le p_T \le 8$~GeV/$c$. We compare
the measured cross section with expectations from Monte Carlo QCD
generators.  The relative yield of $\omega$ to $\pi^0$ mesons is used
to extract the ratio of vector to pseudoscalar meson production for
the first generation of quarks.
\end{abstract}


\narrowtext

\section{INTRODUCTION}

The production of mesons at large transverse momenta ($p_T$) in
hadronic collisions is a process that can be used to study the
phenomenology of Quantum Chromodynamics (QCD) and the parton
fragmentation process. Measurements of the spectra of particles
produced in the fragmentation of partons influence our understanding
of high-$p_T$ processes and affect the design and implementation of
Monte Carlo event generators. Since mesons constitute the majority of
the particles produced in hadronic interactions at high-$p_T$, such
measurements are of interest across a broad spectrum of experiments in
high energy physics.  Although much is known about pion production in
this regime, the hadroproduction of $\omega$ mesons has not been as
extensively studied~\cite{diakonou,donaldson,e629-pi-eta}.  A
comparison of $\omega$ to $\pi^0$ production, which reflects the
overall ratio of vector meson to pseudoscalar meson production
($V/P$), can also be used to sharpen the value of the $V/P$
phenomenological parameter used in current Monte Carlo event
generators.

We report on $\omega$ production in $\pi^-$Be collisions at
515~GeV/$c$, as measured in E706, an experiment which was designed to
study the production of direct photons, neutral mesons, and associated
particles at high-$p_T$ using the Meson West Spectrometer at
Fermilab~\cite{E706-kt}.  The apparatus included a charged particle
spectrometer consisting of silicon microstrip detectors in the target
region and multiwire proportional chambers and straw tube drift
chambers downstream of a large aperture analysis magnet~\cite{charm}.
The target consisted of two 0.8~mm thick copper foils, immediately
upstream of two cylinders of beryllium, one 3.7~cm long and the other
1.1~cm long.  Photons were detected in a 3~m diameter, lead and
liquid-argon, sampling electromagnetic calorimeter (EMLAC), located
$\approx9$~m downstream of the target~\cite{E706-nim}.  The EMLAC
readout was subdivided azimuthally into octants, each consisting of
interleaved, finely segmented, radial and azimuthal views.  The radial
views were also used to form a fast high-$p_T$ event selection
trigger.  Trigger decisions were based on global (full octant) and
local (sixteen 5.5~mm strips) sums of EMLAC energy weighted to measure
transverse momentum~\cite{charm}.

\section{DATA ANALYSIS}

The data sample corresponds to an integrated luminosity of
$8.6$~events/pb.  The measurement of the $\omega$ meson cross section
was based on its decay $\omega\to\pi^0\gamma$. Although this mode has
a branching ratio of only 8.5\%~\cite{pdg}, its decay chain to three
photons yielded a clear signature.  In the following section, we
briefly describe the analysis procedures used to extract the $\omega$
signal in these data.  A more detailed description of the procedures
used to select and reconstruct the events, and to correct the data for
losses due to trigger inefficiencies and selection requirements, can
be found elsewhere~\cite{lucy_thesis,p_pi_eta}.

\subsection{Signal extraction}

\begin{figure}
\epsfxsize=3truein
\centerline{\epsffile{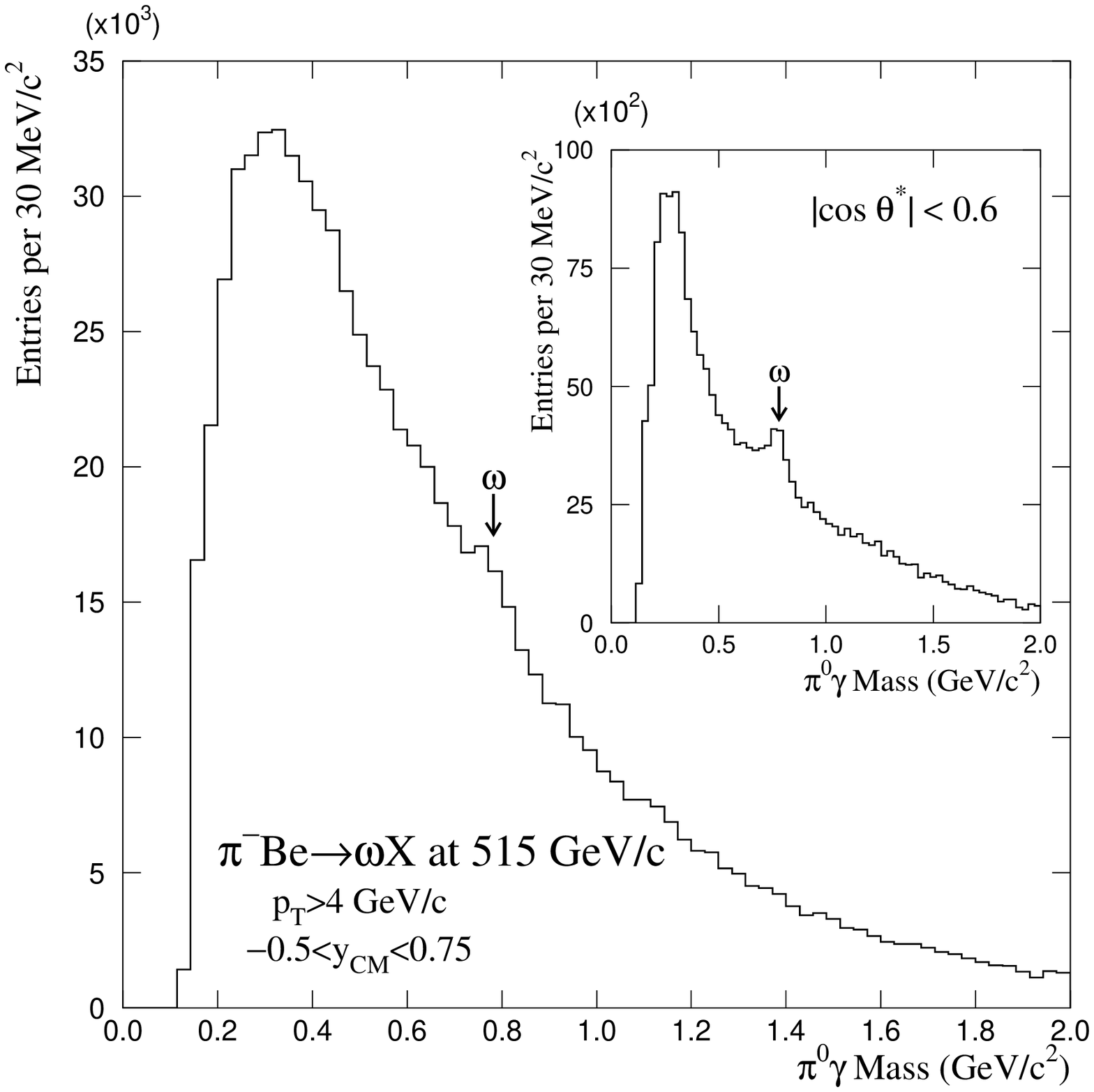}}
\caption{Invariant mass distribution of the $\pi^0\gamma$ system
subject to minimal selection criteria.  The insert shows the
corresponding mass distribution with the additional requirement of
$|\cos\theta^\star|<0.6$.
\label{omegadist}}
\end{figure}

The invariant mass distribution for $\pi^0\gamma$ pairs, subject to
only minimal kinematic cuts, is illustrated in Fig.~\ref{omegadist}.
In our $\omega$ study, we defined a $\pi^0$ as a combination of two
photons with invariant mass, $M_{\gamma\gamma}$, in the range
110~MeV/$c^2 < M_{\gamma\gamma} < 165$~MeV/$c^2$ and energy asymmetry
[$A=\vert E_1 -E_2 \vert/(E_1 + E_2)$] less than 0.75.  The $\omega$
signal in Fig.~\ref{omegadist} consists of a shoulder riding on a
steeply falling background.  To improve the signal-to-background
ratio, we investigated the $\cos\theta^\star$ distribution, where
$\theta^\star$ is defined as the decay angle of the $\pi^0$ in the
$\pi^0\gamma$ rest frame, relative to the $\pi^0\gamma$ line of
flight.  Monte Carlo studies showed that the $\cos\theta^\star$
distribution for accidental pairings of $\pi^0$'s and $\gamma$'s is
peaked near $\pm1$~\cite{lucy_thesis}, whereas the distribution for
unpolarized $\omega$'s is isotropic. (Since we detected no evidence
for $\omega$ polarization in our data, we assumed for the purposes of
the analysis presented in this paper that the signal was dominated by
unpolarized $\omega$'s.)  The requirement
$\vert\cos\theta^\star\vert<0.6$ eliminates a large fraction of the
combinatorial background~\cite{lucy_thesis}, resulting in the
distribution shown in the insert in Fig.~\ref{omegadist}, in which the
$\omega$ signal stands out much more clearly.

\begin{figure}
\epsfxsize=3truein
\centerline{\epsffile{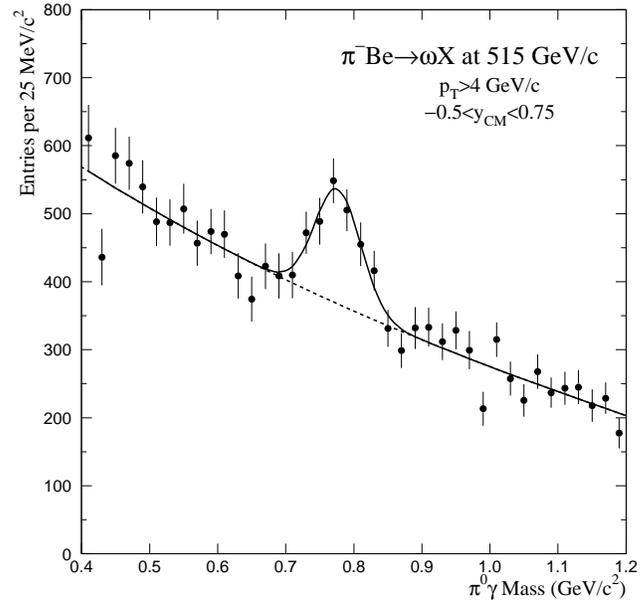}}
\caption{Invariant mass distribution of $\pi^0\gamma$ system 
with the cuts and corrections described in the text.  The background
is parameterized by a third-order polynomial.
\label{omegafit}}
\end{figure}

Figure~\ref{omegafit} displays the weighted mass distribution of
$\pi^0\gamma$ pairs in the vicinity of the $\omega$ mass.  These data
have also been subjected to additional kinematic constraints including
a $\pi^0$ sideband subtraction which Monte Carlo studies showed to be
effective in removing $\eta\gamma$ combinations that peaked in the
vicinity of the $\omega$ signal~\cite{lucy_thesis}.  The effect of
these and other kinematic criteria was estimated using the Monte Carlo
simulation described below.

Fits to the $\pi^0\gamma$ mass spectra for different intervals of
$p_T$ were used to extract the contribution of the $\omega$ signal as
a function of $p_T$.  A Gaussian shape for the signal, combined with a
third order polynomial for the background, yielded a reasonable
description of the data.

\subsection{Monte Carlo}

\begin{figure}
\epsfxsize=3truein
\centerline{\epsffile{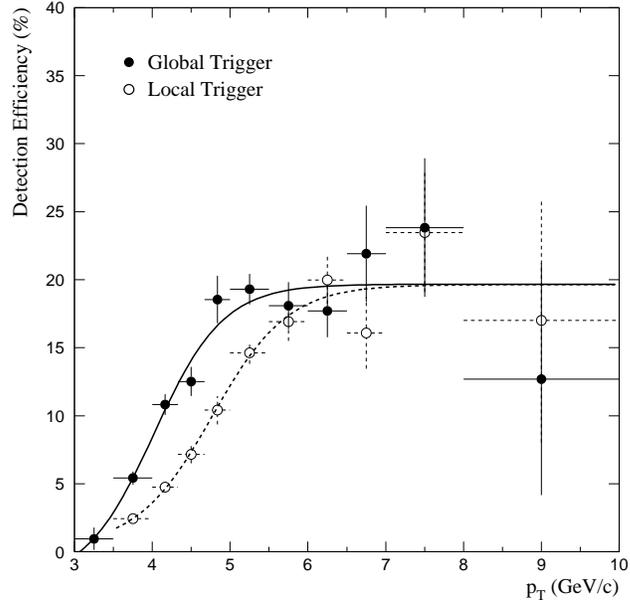}}
\caption{Efficiency for detecting the $\omega$ signal 
in the $\pi^0\gamma$ decay mode, as a function of $p_T$, including
losses due to geometric and trigger acceptances.
\label{eff}}
\end{figure}

A full event simulation was used to evaluate most corrections to the
cross section.  This simulation relied on the {\sc
herwig}~\cite{herwig56} event generator and a {\sc
geant}~\cite{geant3} Monte Carlo simulation of our
apparatus~\cite{p_pi_eta}.  Events containing leading $\omega$ mesons
were generated and reconstructed using our standard reconstruction
package~\cite{E706-nim,p_pi_eta}.  The ratio of reconstructed to
generated events was parameterized as a function of $p_T$, fitted with
a Theta function convolved with a Gaussian (Fig.~\ref{eff}), and
applied to the $\omega$ spectrum.  Due to the relatively large opening
angle for photons in $\omega$ decay, and the resulting spatial spread
of these photons in the EMLAC, the decay products infrequently
satisfied the E706 high-$p_T$ trigger. The overall detection
efficiency for the $\omega$ (in the $\pi^0\gamma$ decay mode) was
$\approx20$\% at high-$p_T$, and differed between the global and local
triggers at lower-$p_T$.

\subsection{Systematic uncertainties}

We estimate the combined systematic uncertainty in the $\omega$ cross
section due to reconstruction efficiency and normalization to be
15\%~\cite{lucy_thesis}.  The energy response of the electromagnetic
calorimeter was calibrated using $\pi^0$, $\eta$, $\omega$, and
$J/\psi$ mass peaks.  The energy scale uncertainty was determined to
be less than 0.5\%~\cite{E706-nim}, and contributed between 5\% and
12\% (as a function of $p_T$) to the systematic uncertainty of the
$\omega$ cross section.

The uncertainty in the fitted background was estimated by using
different parameterizations for the background, and by varying the
$\pi^0\gamma$ mass range and the bin sizes used in the fit. We
estimate a 10\% contribution to the systematic uncertainty in the
determination of the signal due to the fitting procedure.  The
uncertainty in the trigger corrections was estimated by comparing the
$\omega$ cross section obtained using samples selected with different
triggers.  This uncertainty ranged from 17\% for $\omega$ mesons with
$p_T$ of 3.5~GeV/$c$ to 2\% for 8~GeV/$c$.  These quoted systematic
uncertainties do not explicitly incorporate contributions due to the
possibility of $\omega$ polarization, however, as already stated, we
detected no evidence for $\omega$ polarization over the range
$\vert\cos\theta^\star\vert<0.6$.

The systematic uncertainties, combined in quadrature, are quoted with
the cross sections in Table~\ref{omega_table}.  The overall systematic
uncertainty on the $\omega$ cross section was 30\% at
$p_T=$3.5~GeV/$c$, 22\% at 5~GeV/$c$, and 20\% at 8~GeV/$c$.

\section{Results}

\begin{figure}
\epsfxsize=3truein
\centerline{\epsffile{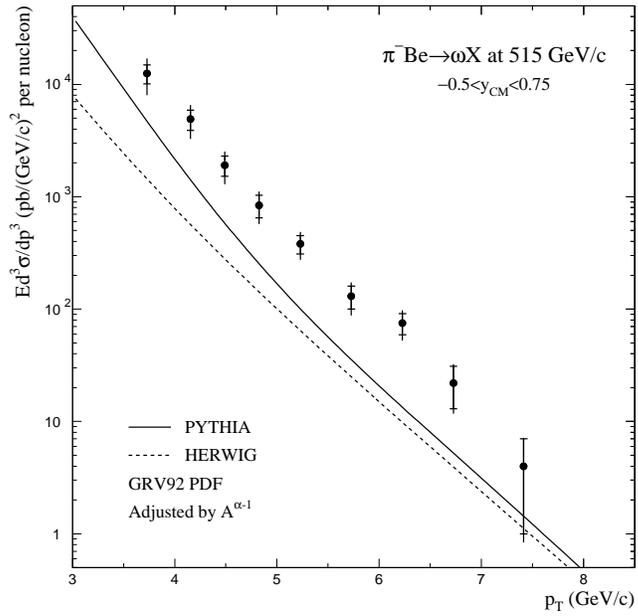}}
\caption{Inclusive invariant differential cross section per nucleon 
for $\omega$ production as a function of $p_T$. The curves correspond
to expectations from Monte Carlo QCD generators, and have been
adjusted for nuclear effects.
\label{ptomega}}
\end{figure}

Table~\ref{omega_table} lists our measured inclusive invariant cross
section for $\omega$ meson production in $\pi^-$Be interactions at
515~GeV/$c$ along with statistical and systematic uncertainties.  The
results are binned in $p_T$ from 3.5 to 8~GeV/$c$, and averaged over
the range of our acceptance in rapidity ($-0.5 < y_{\rm CM} < 0.75$).
The $\omega$ cross section is also displayed in Figure~\ref{ptomega},
and compared with expectations from {\sc pythia}~\cite{pythia61} and
{\sc herwig}~\cite{herwig61}.  The predictions from both Monte Carlos
are substantially smaller than the measured $\omega$ cross section.
Comparison between the relative yields of $\omega$'s originating in
the Be and Cu target materials resulted in the value
$\alpha=1.12\pm0.07\pm0.07$~\cite{lucy_thesis} using the
parameterization $\sigma_A \propto A^\alpha$.  The Monte Carlo results
have been adjusted for this nuclear effect.

\begin{figure}
\epsfxsize=3truein
\centerline{\epsffile{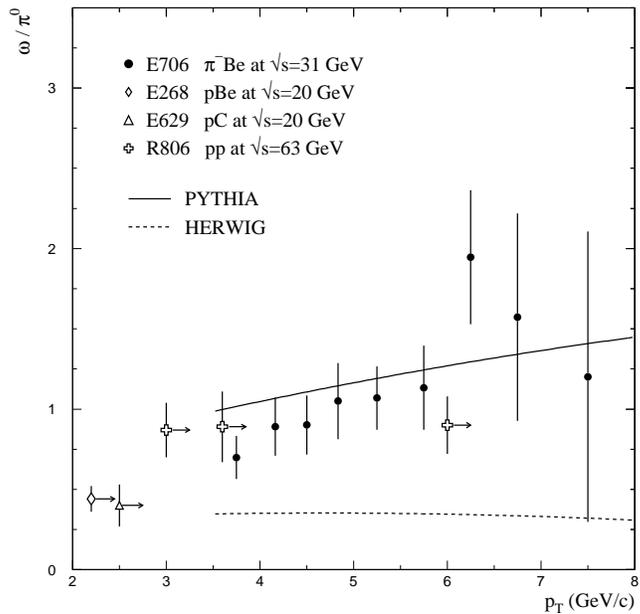}}
\caption{Ratio of the invariant cross section for $\omega$ production
to the $\pi^0$ cross section as a function of $p_T$.  Results of this
experiment are compared with previous measurements, and with
expectations from {\sc pythia} and {\sc herwig} (assuming $\omega$ and
$\pi^0$ production have the same nuclear dependence).  Results from
earlier experiments are integrated over $p_T$, and displayed at their
minimum $p_T$ values.
\label{omegapi}}
\end{figure}

Figure~\ref{omegapi} and Table~\ref{omega_table} display the relative
yields of $\omega$ and $\pi^0$ mesons measured in E706, in terms of
the ratio of the inclusive differential cross sections as a function
of $p_T$.  The prediction from the {\sc pythia}~\cite{pythia61}
generator is consistent with our measured ratio.  The ratio from {\sc
herwig}~\cite{herwig61} is much smaller than both our measurement and
the result from {\sc pythia}.  We include for comparison three
previous results on the $\omega$ to $\pi^0$ cross section ratio for
incident protons on p, Be, and C
targets~\cite{diakonou,donaldson,e629-pi-eta}. These earlier
measurements were integrated over $p_T$ and displayed at their minimum
$p_T$ value.

The $\omega$ to $\pi^0$ ratio can be used to determine the value of
$V/P$.  The quark content of both $\omega$ and $\pi^0$ is the same,
rendering their production ratio insensitive to beam and target
composition.  Corrections to $V/P$ to account for indirect production
were determined using both {\sc pythia}~\cite{pythia61} and {\sc
herwig}~\cite{herwig61}. In both cases, the data from E706 were found
to require only a relatively small correction for indirect
contributions from the decay of higher mass hadrons.  We note,
however, that in {\sc pythia} almost all $\omega$ mesons are produced
directly, which seems to us an extreme assumption, and the {\sc
herwig} $\omega$ to $\pi^0$ ratio differs greatly from our
data. Nevertheless, if we use these Monte Carlos to correct our
measured $\omega$ to $\pi^0$ ratio for indirect production, the
resulting $V/P$ values are $1.2\pm0.1$ using {\sc pythia} and
$0.9\pm0.1$ using {\sc herwig}.

\acknowledgments

We thank the management and staff of Fermilab, the U.~S. Department of
Energy, the National Science Foundation, including its Office of
International Programs, and the Universities Grants Commission of
India, for their support of this research. We are also pleased to
acknowledge the contributions of our colleagues on Fermilab experiment
E672 to this and other aspects of E706.

%%%%%%%%%%%%%%%%%%%%%%%%%%%%%%%%%%%%%%%%%%%%%%%%%%%%%%%%%%%%%%%%%%%
\renewcommand{\baselinestretch}{1.}
\newpage
\bibliography{omega}  
\bibliographystyle{prsty}
%%%%%%%%%%%%%%%%%%%%%%%%%%%%%%%%%%%%%%%%%%%%%%%%%%%%%%%%%%%%%%%%%%%
%
% Figures
%
%%%%%%%%%%%%%%%%%%%%%%%%%%%%%%%%%%%%%%%%%%%%%%%%%%%%%%%%%%%%%%%%%%%
\widetext
\newpage
\newpage
\newpage
\newpage
\newpage
%%%%%%%%%%%%%%%%%%%%%%%%%%%%%%%%%%%%%%%%%%%%%%%%%%%%%%%%%%%%%%%%%%%
%
% Tables
%
%%%%%%%%%%%%%%%%%%%%%%%%%%%%%%%%%%%%%%%%%%%%%%%%%%%%%%%%%%%%%%%%%%%
\newpage
\narrowtext
\begin{table}
\caption{Invariant differential cross section per nucleon
$\left( Ed^3\sigma/dp^3 \right)$ for the inclusive reaction $\pi^- +
{\rm Be} \rightarrow \omega + X$, averaged over the rapidity range,
$-0.5 < y_{\rm CM} < 0.75$.  Also included is a ratio of the $\omega$
cross section to the corresponding $\pi^0$ cross section measured by
E706. }
\begin{tabular}{ccr@{$\pm$}r@{$\pm$}rr@{$\pm$}r@{$\pm$}r}
$p_T$ & $\langle p_T\rangle$ &\multicolumn{3}{c}{$Ed^3\sigma/dp^3$} 
      &\multicolumn{3}{c}{$\omega/\pi^0$}\\
(GeV/$c$) &(GeV/$c$)&\multicolumn{3}{c}{$pb/($GeV$/c)^{2}$}
          &\multicolumn{3}{c}{ratio} \\
\tableline
{$3.5 - 4.0$} & 3.73 & 12500 &  2400 & 3800 & 0.70 & 0.13 & 0.20\\ 
{$4.0 - 4.3$} & 4.16 &  4900 &  1000 & 1300 & 0.89 & 0.18 & 0.22\\ 
{$4.3 - 4.7$} & 4.49 &  1910 &   390 &  480 & 0.90 & 0.18 & 0.20\\ 
{$4.7 - 5.0$} & 4.83 &   840 &   190 &  190 & 1.05 & 0.24 & 0.21\\ 
{$5.0 - 5.5$} & 5.23 &   380 &    70 &   80 & 1.07 & 0.20 & 0.20\\ 
{$5.5 - 6.0$} & 5.73 &   130 &    30 &   30 & 1.1  & 0.3  & 0.2\\ 
{$6.0 - 6.5$} & 6.23 &    75 &    16 &   16 & 2.0  & 0.4  & 0.4\\ 
{$6.5 - 7.0$} & 6.73 &    22 &     9 &    5 & 1.6  & 0.6  & 0.3\\ 
{$7.0 - 8.0$} & 7.41 &     4 &     3 &    1 & 1.2  & 0.9  & 0.3\\ 
\end{tabular}
\label{omega_table}
\end{table}
\vfil
\end{document}